\documentclass[natbib]{svmult}

\usepackage{bm}
\usepackage{amsmath}
\usepackage{amssymb}
\usepackage{makeidx}     
\usepackage{graphicx}    
\usepackage{multicol}    

\makeindex             

\begin{document}

\title*{Modeling Stochastic Clonal Interference}
\author{Paulo R.A. Campos\inst{1}\and
Christoph Adami\inst{1,}\inst{2}\and
Claus O. Wilke\inst{1}}
\institute{Digital Life Laboratory 136-93, California Institute of Technology, Pasadena, CA 91125
\and Jet Propulsion Laboratory 126-347, California Institute of Technology,
Pasadena, CA 91109}
%
%
\maketitle

\begin{abstract}
  We study the competition between several advantageous mutants in an asexual
  population (clonal interference) as a function of the time between the appearance
   of the mutants $\Delta t$, their selective advantages, and the rate of deleterious mutations.
  We find that the overall probability of fixation (the probability that at least
  one of the mutants becomes the ancestor of the entire population) does not depend
  on the time interval between the appearance of these mutants, and
  equals the probability that a genotype bearing all of these mutations reaches fixation.
  This result holds also in the presence of deleterious mutations, and for an
  arbitrary number of competing mutants. We also show that if mutations interfere,
  an increase in the mean number of fixation events is associated with a decrease in the expected
  fitness gain of the population.
  \end{abstract}

\section{Introduction}

Evolution, according to~\citet{Dobzhansky1973}, is the unifying concept that pulls
together all the different strands of biology. Indeed, while evolution provides the
framework to understand the otherwise bewildering panoply of adapted forms,  it
also allows us to understand the patterns that mutation and selection leave in the
molecules of life, namely DNA and proteins. One of the central problems of the
branch of biology that is devoted to the systematics of the tree of all living
things, molecular evolution, concerns the rate of adaptation of individual
organisms and species. A precise understanding of molecular phylogeny requires
accurate models
of evolution and adaptation. In this contribution, we model the adaptation of
chromosomes that do not undergo recombination, or more generally, the rate of
adaptation of asexual organisms.

The main observation that we study here is that the rate of adaptation of asexual
organisms, or of regions of low recombination in the genomes of sexual organisms,
does not steadily increase with increasing rate of advantageous mutations. When two
or more advantageous mutations appear in different organisms at approximately the
same time, then only one of them can go to fixation, i.e., become shared by all
members of the population, while the others will be lost. This loss of potentially
beneficial mutations limits the rate of adaptation to the rate at which individual
mutants can go to fixation.  By contrast, in sexual organisms, several mutations
can recombine and thus go to fixation together. This interference of advantageous
mutations has long been recognized as a potential disadvantage of asexual
populations \citep{fisher30,Muller64,hill66}. Recently, several groups have worked
on an exact quantification of the interference effect, both in theoretical studies
\citep{barton95,gerrish98,orr2000,gerrish2001,McVeanCharlesworth2000} and in
experimental studies with bacteria \citep{visser99,rozen2002,shaver2002} and
viruses \citep{miralles99,cuevas2002}. A good quantitative understanding of the
interference effect is necessary in order to assess the influence that interference
has on the patterns of molecular evolution and variation in large populations.

There are two separate dynamics that both contribute to the overall interference
effect: First, if two advantageous mutants are both present in sufficiently high
concentrations, such that loss to drift can be neglected, then they compete
deterministically, and the mutant with the higher selective advantage will replace
the other one. Second, if at least one advantageous mutant is still very rare, then
we have to consider the influence of other mutants' presence on the chance that
this mutant is lost to drift. To date, there is no single theory that takes into
account both dynamics to their full extent. \citet{gerrish98} calculated the speed
of adaptation as a function of population size and beneficial mutation rate, under
the assumption that clonal interference can be neglected during the initial phase
of drift. They assumed that the probability that a mutant is not lost to drift
corresponds to one minus the standard probability of fixation (as calculated for
example by \citealt{fisher22}, \citealt{haldane27}, \citealt{kimura62}) of that
mutant. \citet{orr2000} modified the calculations of Gerrish and Lenski to include
beneficial mutations that arise in genomes bearing one or more deleterious
mutations, but also did not address the effect of interference of other beneficial
mutations during the initial phase of drift. In general, the influence of
deleterious mutations on the probability of fixation has been studied extensively
\citep{manning84,charlesworth94,peck94,johnson2002}, but there are very few studies
that consider the effect of interfering beneficial mutations. The problem of
interfering advantageous mutations is that there exists no simple theoretical
framework with which their influence on drift can be described accurately.
\citet{barton95} derived an approximation that allowed him to calculate the
probability of fixation of a mutant in a population that is undergoing a selective
sweep. However, his approximation is valid only for very small selective
advantages. \citet{McVeanCharlesworth2000} studied a similar situation with
numerical simulations, and applied their results to codon bias and levels of
polymorphism in molecular evolution.

Here, we study the mutual interference of two advantageous mutants that are each
initially present in only a single organism. We derive a phenomenological
description of the mutants' fixation probabilities in the interference regime. This
phenomenological description, which is in essence an interpolation between limiting
cases that can be described with standard theory, agrees very well with numerical
simulations. We also find that the expected fitness increase of the population is
maximized if the mutant with higher selective advantage arrives earlier, even
though this order of appearance leads, on average, to fewer mutants that go to
fixation.

\section{Model}

We consider three distinct sequence types: wild type, advantageous mutant 1,
and advantageous mutant 2, with fitness values 1, $1+s_1$, and $1+s_2$,
respectively.  Thus, $s_{1}$ denotes the selective advantage of mutants of
type $1$, and $s_{2}$ is the selective advantage of mutants of type $2$. For
simplicity, we assume that $s_{2} > s_{1}$. Initially, the population is
homogeneous and consists of the wild type only. Then, one randomly chosen wild
type individual is replaced by an advantageous mutant of either type, and some
$\Delta t$ generations later, another randomly chosen individual of the
population (either wild type or mutant) is replaced by a mutant of the other
type. Throughout this paper, we will understand $\Delta t$ to be the
difference in generations between the appearance of mutant 2 and mutant 1, so
that negative values of $\Delta t$ indicate that mutant 2 appeared before
mutant 1.

The population is finite of size $N$, and replication takes place
in discrete generations, according to the Wright-Fisher model: All individuals
in generation $t$ are direct descendants of the individuals of the previous
generation; the probability that an individual is the offspring of a
particular parent is proportional to the parent's fitness. We
introduce deleterious mutations into the offspring organisms with probability
$u$. If a mutant of type 1 or 2 is hit by a mutation, then its fitness is set
to 1 (that is, it reverts to the lower fitness wild type). We do not consider
deleterious mutations in the wild type genotype.

We consider a genotype to be fixed if it has become the most-recent common ancestor
of the whole population, regardless of whether some individuals in the population
have a different genotype. This definition of fixation has been used recently to
study fixation of beneficial mutations in a heterogeneous genetic background
\citep{barton95,johnson2002}, and also to investigate the process of fixation in a
viral quasispecies \citep{wilke2003}.

\section{Theoretical analysis}

\subsection{Probability of ultimate fixation}

In the Appendix, we derive equations for the probability of fixation $P(s,u)$
and time to fixation $T(s,u)$ of an individual mutant with selective advantage
$s$ and mutation rate $u$. For two mutants, we are mostly
interested in the probability of ultimate fixation, that is, the probability
that a mutant reaches fixation and is not subsequently replaced by the other
mutant. In the following, we will denote the probability that mutant $i$
reaches ultimate fixation under the condition that the two mutants are
introduced $\Delta t$ generations apart as $\pi_i(\Delta t)$. We can calculate
$\pi_i(\Delta t)$ for the two limiting cases $\Delta t\rightarrow -\infty$ and
$\Delta t\rightarrow \infty$, and present a phenomenological description for
intermediate $\Delta t$.

For $\Delta t\rightarrow -\infty$, that is, when mutant 2 arises much earlier
than mutant 1, we have
\begin{align}\label{p1minusinf}
\pi_1(-\infty) &= [1-P(s_2,u)]P(s_1,u)\,,\\\label{p2minusinf}
\pi_2(-\infty) &= P(s_2,u)\,.
\end{align}
Since $s_2>s_1$, mutant 1 can reach fixation only when mutant 2 has not
reached fixation. For $\Delta t\rightarrow \infty$, we have to consider the possibility
that mutant 1 reaches fixation first, but is later replaced by mutant
2. Therefore, we find
\begin{align}\label{p1inf}
\pi_1(\infty) &= P(s_1,u)\Big[1-P\Big(\frac{1+s_2}{(1+s_1)(1-u)}-1,u\Big)\Big]\,,\\\label{p2inf}
\pi_2(\infty) &= P(s_1,u)P\Big(\frac{1+s_2}{(1+s_1)(1-u)}-1,u\Big)+[1-P(s_1,u)]P(s_2,u)\,,
\end{align}
where $P(\frac{1+s_2}{(1+s_1)(1-u)}-1,u)$ is the probability that mutant
2 goes to fixation after mutant 1 has already reached fixation. Since
$P(\frac{1+s_2}{(1+s_1)(1-u)}-1,u)$ is always smaller than or equal to
$P(s_2,u)$, we have $\pi_1(-\infty)\leq \pi_1(\infty)$ and $\pi_2(-\infty)\geq
\pi_2(\infty)$. In other words, both mutants have a higher probability of
fixation when they are introduced first than when they are introduced second.

For intermediate values of $\Delta t$, there is no theory that enables us to derive
a simple expression for $\pi_i(\Delta t)$ (but see \citealt{barton95}). We cannot
use branching process theory, because it assumes that the presence of the invading
mutants does not influence the mean fitness (which they do for intermediate $\Delta
t$). Also, diffusion theory becomes unwieldy when there are more than 2 different
sequence types. Nevertheless, we have found that we can develop a phenomenological
description of the competition of two mutants as follows. We know that $\pi_i(t)$
must reach the two limiting values $\pi_i(\infty)$ and $\pi_i(-\infty)$ for
sufficiently large positive or negative $\Delta t$. Moreover, as long as $\Delta
t<0$, we do not expect $\pi_i(\Delta t)$ to be very different from
$\pi_i(-\infty)$, because mutant 1 has only a realistic chance of proliferating and
going to fixation if mutant 2 is not present in the population. As long as $\Delta
t\lesssim0$, mutant 2 will either go to fixation relatively unscathed from the
later appearance of mutant 1, or it will be lost to drift, in which case mutant 1
will have its turn.  Likewise, for $\Delta t$ larger than the time to fixation of
mutant 1, $T_1$, we expect that $\pi_i(\Delta t)=\pi_i(\infty)$, because $T_1$
generations after the introduction of mutant 1, mutant 2 will either find a
population in which mutant 1 has already gone to fixation, or one in which it has
been lost to drift. For $0\lesssim\Delta t\lesssim T_1$, we expect that
$\pi_1(\Delta t)$ smoothly increases from $\pi_1(-\infty)$ to $\pi_1(\infty)$,
while $\pi_2(\Delta t)$ smoothly decreases from $\pi_2(-\infty)$ to
$\pi_2(\infty)$.  These considerations suggest a sigmoidal form for $\pi_i(\Delta
t)$. We use a logistic growth model to describe the smooth transition from
$\pi_i(-\infty)$ to $\pi_i(\infty)$ within the range $0\lesssim\Delta t\lesssim
T_1$:
\begin{equation}\label{tanh-ansatz}
 \pi_i(\Delta t)=\pi_i(-\infty) + \frac{\pi_i(\infty)-\pi_i(-\infty)}
    {1+e^{-\gamma_1\left(\Delta
t-T_1/2\right)}}\;,
\end{equation}
where $\gamma_1$ is the fitness advantage of mutant 1 at finite mutation rate (see
the definition following Eq. (\ref{pieq}) below). This expression appears to be an
acceptable description of the exact time dependence (see Numerical Results) without
free parameters.

\subsection{Overall probability of fixation}

Besides the individual probabilities $\pi_1(\Delta t)$ and $\pi_2(\Delta t)$,
their sum $\pi=\pi_1(\Delta t)+\pi_2(\Delta t)$ is also of
interest. This sum is the overall probability that at least one mutant goes to
fixation. When we sum Eqs.~(\ref{p1minusinf}) and~(\ref{p2minusinf}), or
Eqs.~(\ref{p1inf}) and~(\ref{p2inf}), we find that the overall fixation
probability does not depend on the order in which the mutants are introduced,
and has the form
\begin{equation}\label{overall-fix-prob}
  \pi = P(s_1,u)+P(s_2,u)-P(s_1,u)P(s_2,u)\,.
\end{equation}
Furthermore, if Eq.~(\ref{tanh-ansatz}) is indeed a good description of
$\pi_i(\Delta t)$ for arbitrary $\Delta t$, then $\pi$ should not depend on
$\Delta t$ at all, because all time dependencies cancel when we use
Eq.~(\ref{tanh-ansatz}) to calculate $\pi_1(\Delta t) + \pi_2(\Delta t)$.

The invariance of the overall fixation probability $\pi$ under the order by
which the mutants appear is a general property. It holds also when more than 2
mutants arise: Assume that $n$ advantageous mutants arise, with selective
advantages $s_1,\dots,s_n$ compared to the wild type. Further assume that the
time intervals between the appearances of the mutants are large. Then, the
probability that none of the mutants make it to fixation is
$\prod_{i=1}^n[1-P(s_i,u)]$, regardless of the order of their appearance. The
probability that at least one mutant goes to fixation is therefore
\begin{equation}\label{overall-fix-prob-multi}
  \pi = 1-\prod_{i=1}^n[1-P(s_i,u)]\,,
\end{equation}
which reduces to Eq.~(\ref{overall-fix-prob}) in the case of $n=2$.
Using an extension of Kimura's well-known result for $P(s_i,u)$ derived in the Appendix,
Eq.~(\ref{kimura-approx}), we can simplify this expression
even further.  We find  in the limit $N\rightarrow \infty$:
\begin{equation}\label{pieq}
\pi = 1- \exp\left[-2\left(\gamma_{1}+\gamma_{2}+ \cdots +\gamma_{n}\right)\right]\,,
\end{equation}
with $\gamma_i=(1+s_i)(1-u)-1$ for $u<s_i/(1+s_i)$, and $\gamma_i=0$
otherwise. Moreover, in the absence of deleterious mutations, for $u=0$, $\pi$
becomes
\begin{equation}
\pi = 1 - \exp\left[-2\left(s_{1}+s_{2}+ \cdots +s_{n}\right)\right]\,.
\end{equation}
In this limit, the overall probability of fixation is the same as the probability
of fixation of a single mutant with selective advantage $s$ equal to the sum of the
selective advantages of all invading mutants, $s=\sum_i s_{i}$. Such a hypothetical
mutant is extremely unlikely in clonal populations, because all beneficial
mutations would have to hit the lineage sequentially, but could occur when
beneficial mutations are shared via recombination.  In the limit that all
$\gamma_i$ vanish (that is, for very large mutation rates or for vanishing $s_i$),
Eq.~(\ref{overall-fix-prob-multi}) becomes
\begin{equation}
  \pi = \frac{n}{N}\,. \label{neutral}
\end{equation}

\subsection{Expected fitness increase and expected number of fixed mutants}

Depending on which mutant goes to fixation, the average fitness of the final
population is either 1, $1+\gamma_1$, or $1+\gamma_2$, with $\gamma$ as
defined following Eq.~(\ref{pieq}). The expected fitness
increase $\langle \gamma(\Delta t)\rangle$ after the introduction of the two
mutants is therefore
\begin{equation}\label{expectedinc}
  \langle \gamma(\Delta t)\rangle = \gamma_1 \pi_1(\Delta t) + \gamma_2
  \pi_2(\Delta t)\,.
\end{equation}
Since we know that $\pi_1(\Delta t)+\pi_2(\Delta t)$ is constant, and
$\gamma_2 \geq \gamma_1$ as a direct consequence of our assumption $s_2>s_1$,
it follows that $\langle\gamma(-\infty)\rangle \geq \langle
\gamma(\infty)\rangle$. If we write $\pi_1(\infty)=\pi_1(-\infty)+\Delta
\pi$, and $\pi_2(-\infty)=\pi_2(\infty)+\Delta \pi$, then we find
\begin{equation}
  \langle \gamma(-\infty)\rangle - \langle \gamma(\infty)\rangle =
    (\gamma_2-\gamma_1)\Delta \pi\geq 0\,.
\end{equation}
Thus, the expected fitness increase is larger if we introduce the mutant with the
higher selective advantage first.

Let $n_{\rm fix}$ denote the number of fixed mutants, that is, we have $n_{\rm
fix}=0$ if none of the mutants reach fixation, $n_{\rm fix}=1$ if exactly one of
the mutants reaches fixation, and $n_{\rm fix}=2$ if first mutant 1 reaches
fixation and is later replaced by mutant 2. For $\Delta t\rightarrow-\infty$ and
$N$ large, $n_{\rm fix}$ can never be larger than one, because the probability that
mutant 1 goes to fixation in the background of mutant 2 is zero. We find in this
limit for the expected value of $n_{\rm
  fix}$:
\begin{equation}
  \langle n_{\rm fix}(-\infty)\rangle = \pi_1(-\infty) + \pi_2(-\infty) = \pi\,.
\end{equation}
In the limit $\Delta t\rightarrow \infty$, on the other hand, we find
\begin{align}
  \langle n_{\rm fix}(\infty)\rangle &= \pi_1(-\infty) + \pi_2(-\infty)
+P(s_1,u)P\Big(\frac{1+s_2}{(1+s_1)(1-u)}-1,u\Big)\notag \\
& = \pi + P(s_1,u)P\Big(\frac{1+s_2}{(1+s_1)(1-u)}-1,u\Big)\,.
\end{align}
Clearly, $\langle n_{\rm fix}(\infty)\rangle\geq\langle n_{\rm
fix}(-\infty)\rangle$. This means that if the mutant with the smaller selective
advantage appears before the mutant with the larger selective advantage, then the
expected number of mutants that go to fixation is larger than if the mutant with
the larger selective advantage appears first. However, at the same time the
expected increase in average fitness is smaller, as we saw in the previous
paragraph.

\section{Numerical Simulation}
In order to measure fixation probabilities, we carried out $100,000$
replicates of the simulation for each set of parameters, and recorded the
final outcome (all individuals unmarked, or marked as descendants of either
advantageous mutant). We studied a population of size $N=1000$ and mutation
rates $u=0.0$, 0.01, 0.02, 0.03, 0.05, 0.07, 0.1, 0.15, 0.20, 0.3, 0.4, 0.5,
0.6, 0.7, 0.8. The selective advantages were $s_{1}=0.1$ and $s_{2}=0.2$,
$s_{1}=0.1$ and $s_{2}=0.5$, $s_{1}=0.05$ and $s_{2}=0.2$. We also
studied a population of size $N=10,000$ for a subset of these parameters, in order
to make sure that our results were robust against a change in population
size. Because of the sizable amount of CPU time needed to carry out 100,000
replicates for $N=10,000$, we could however not study this case exhaustively.

\begin{figure}[ht]
\centering
\includegraphics[width=8cm,angle=0,clip=true]{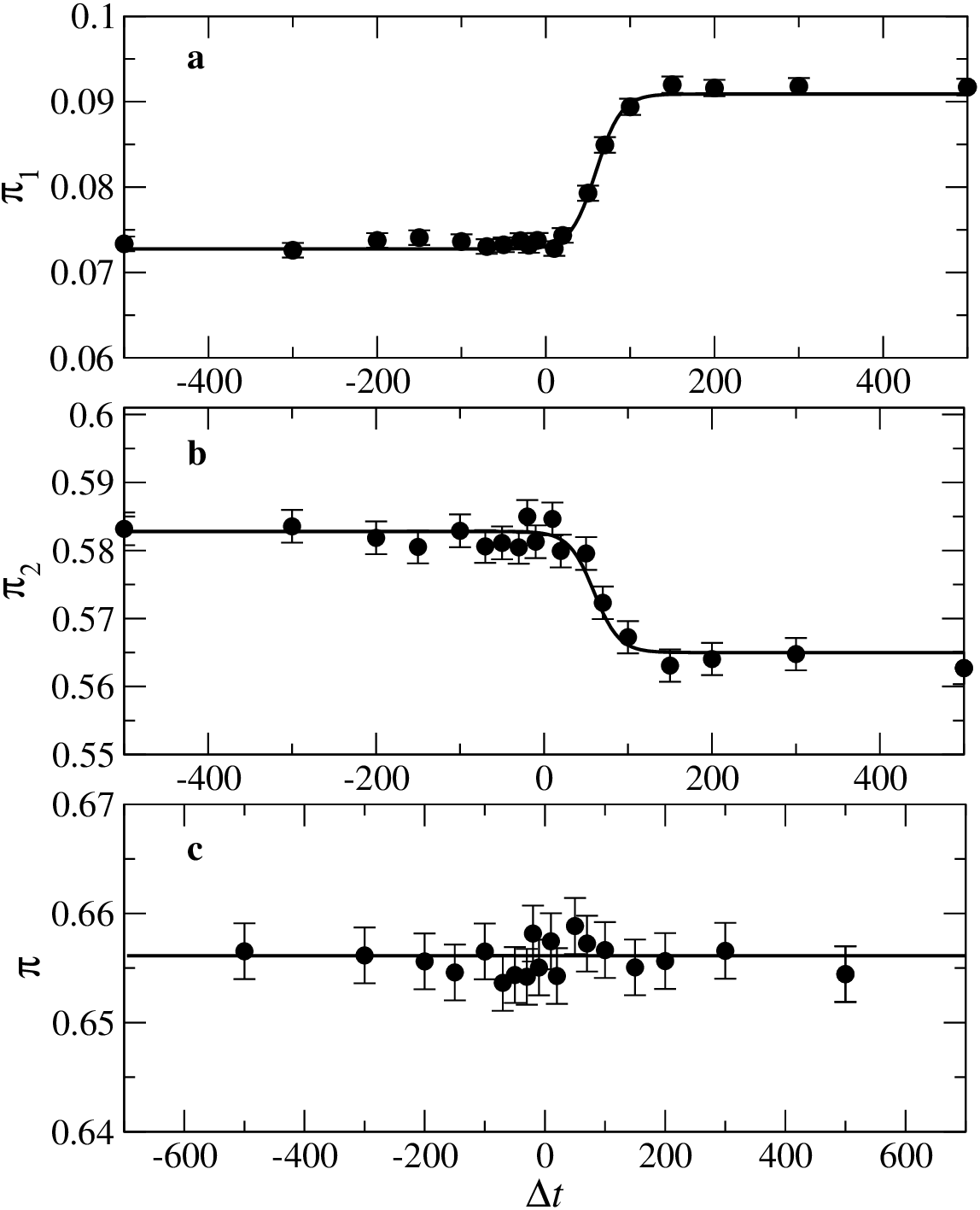}
\caption{\label{figure:fig1} Probability of fixation as a function of time
  interval $\Delta t$ ($N=1000$, $s_{1}=0.1$, $s_{2}=0.5$, $u=0.0$). Solid lines represent
the prediction according to the logistic growth model Eq. (\ref{tanh-ansatz}). a:
Probability of fixation of genotype $1$ (less beneficial mutation), $\pi_{1}$. b:
Probability of fixation of genotype $2$ (more beneficial mutation), $\pi_{2}$. c:
Overall probability of fixation, $\pi$.}
\end{figure}

In order to keep track of the evolutionary history of each mutant, we marked
the initial mutants 1 and 2 with two distinct inheritable neutral markers. In
that way, we could distinguish wild type sequences that were descendants from
mutants 1 or 2 from the wild type sequences that were originally present. We
continued all simulations until all individuals in the population were either
unmarked, marked as descendants of mutant 1, or marked as descendants of
mutant 2.

Figure \ref{figure:fig1} shows the probability of ultimate fixation of mutants 1
and 2, $\pi_1$ and $\pi_2$, and the overall probability of fixation
$\pi=\pi_1+\pi_2$, for $u=0$.  We see that the probability
 $\pi_1$ is constant for $\Delta t<0$, and starts to increase as soon as
$\Delta t$ turns positive. Eventually, $\pi_1$ levels off again. We observe
the opposite behavior for the probability $\pi_{2}$. For $\Delta
t<0$, $\pi_{2}$ is constant, but decreases rapidly in the same range of
$\Delta t$ in which $\pi_{1}$ increases. Finally, $\pi_2$ levels off as well.

\begin{figure}[tbc]
\centering
\includegraphics[width=8cm,angle=0,clip=true]{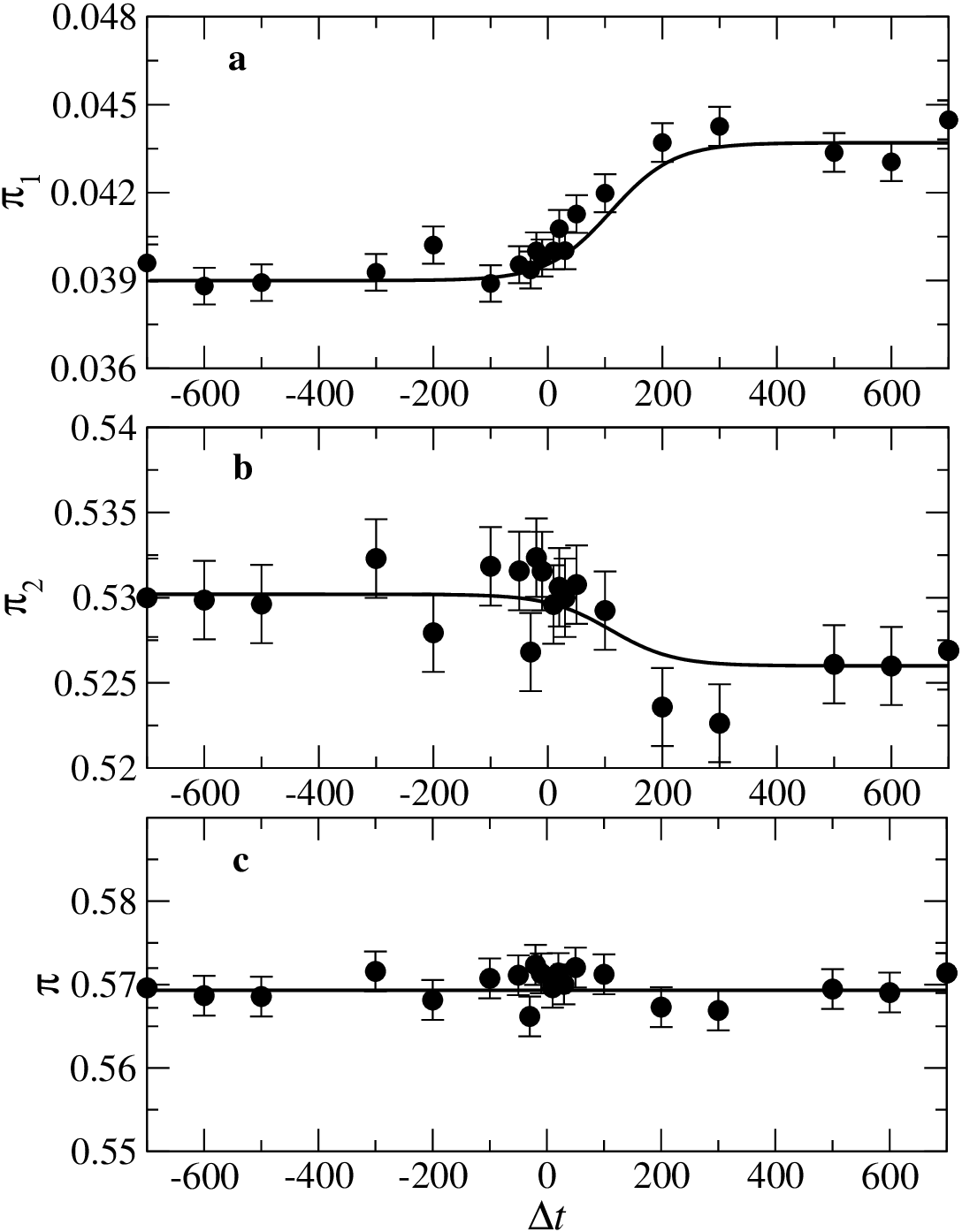}
\caption{\label{figure:fig2} Probability of fixation as a function of time interval
$\Delta t$ ($N=1000$, $s_{1}=0.1$, $s_{2}=0.5$, $u=0.05$). Solid lines represent
the prediction according to the logistic growth model Eq. (\ref{tanh-ansatz}). a:
Probability of fixation of genotype $1$, $\pi_{1}$. b: Probability of fixation of
genotype $2$, $\pi_{2}$. c: Overall probability of fixation, $\pi$.}
\end{figure}

The overall probability of fixation
$\pi$ is approximately constant for all values of $\Delta t$. The solid lines
in Fig.~\ref{figure:fig1}a and~b correspond to Eq.~(\ref{tanh-ansatz}), and
the solid line in Fig.~\ref{figure:fig1}c is
$\pi_1(-\infty)+\pi_2(-\infty)$. We find that our phenomenological description
Eq.~(\ref{tanh-ansatz}) performs very well for intermediate $\Delta t$.

In Fig. \ref{figure:fig2} we plot the same quantities as those shown in Figure $1$,
but now with a positive mutation rate $u=0.05$. We observe that the two
probabilities $\pi_{1}$ and $\pi_{2}$ have smaller values than in the absence of
mutations. The phenomenological description still works well, and appears to
correctly take into account the effects of mutation. The overall fixation
probability $\pi$ is again independent of $\Delta t$.
\begin{figure}[tbc]
\centering
\includegraphics[width=9cm,angle=0,clip=true]{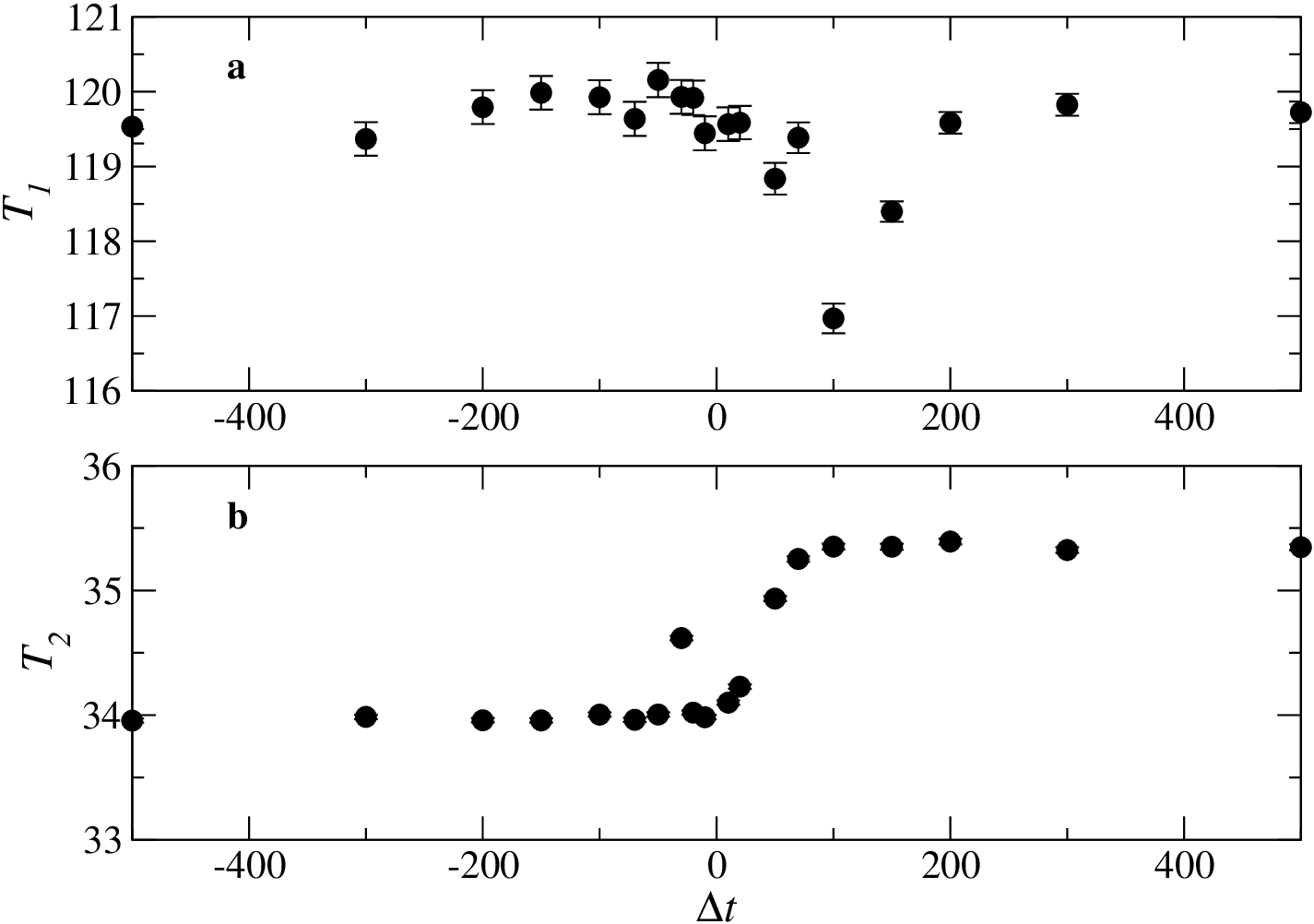}
\caption{\label{figure:figtime} Time to fixation as a function of the time
  interval between beneficial mutants $\Delta t$ ($N=1000$, $s_{1}=0.1$, $s_{2}=0.5$, $u=0.0$).
  a: Time to fixation of genotype 1, $T_{1}$. b: Time to fixation of genotype 2, $T_{2}$.}
\end{figure}
In Figure \ref{figure:figtime} we show the time to fixation of mutant $1$, $T_{1}$,
and the time to fixation of mutant $2$, $T_{2}$, as functions of $\Delta t$.  The
parameter values are the same as those of Figure \ref{figure:fig1}. In Figure
\ref{figure:figtime}a, we see that $T_{1}$ is approximately constant for all values
of $\Delta t$ except for the range $0 \leq \Delta t \leq 200$.  For negative
$\Delta t$ the fixation of mutant 1 occurs only when mutant 2 has been eliminated.
Therefore, $T_{1}$ corresponds to the result for the fixation time of a mutant with
selective advantage $s_{1}$ in a homogeneous population with wild-type individuals
only.  The same is true for large positive $\Delta t$ because there mutant 1 has
enough time to reach fixation without the interference of the second mutant. For
small positive $\Delta t$, we observe a decrease of $T_{1}$. The decrease occurs
for those $\Delta t$ for which the two beneficial mutants coexist for several
generations in the population. Fixation of mutant 1 in this regime occurs only when
mutant 1 reaches fixation so quickly that mutant 2 has not had time to build up
momentum. Otherwise, most likely mutant 1 will be displaced by mutant 2 before
reaching fixation. For mutant 2, the time to fixation $T_2$ is shorter for negative
$\Delta t$ than for positive $\Delta t$. For positive $\Delta t$, in a fraction of
cases mutant 2 has to go to fixation in a background of mutant 1, rather than in a
background of wild type. In the background of mutant 1, the selective advantage of
mutant 2 is smaller than in a wild-type population, which explains the increased
time to fixation. Interestingly, before $T_2$ starts to rise for increasing
positive $\Delta t$, it quickly spikes at $\Delta t\approx 30$. This spike has the
following explanation: If mutant 1 is introduced right before mutant 2 has reached
fixation, then the time to fixation of mutant 2 is increased by the additional time
it takes for mutant 1 to disappear again. This additional time will typically be
one or two generations, which is in agreement with the height of the spike. ($T_1$
on the top of the spike is actually approximately half a generation larger than
before the spike.) The limiting values for $\pi_{1}$ and $\pi_{2}$ when  $\Delta t
\rightarrow -\infty$ and $\Delta t\rightarrow -\infty$ were estimated by means of
branching process theory which provides very good accuracy, especially for low
mutation values.
\begin{figure}[tbc]
\centering
\includegraphics[width=9cm,angle=0,clip=true]{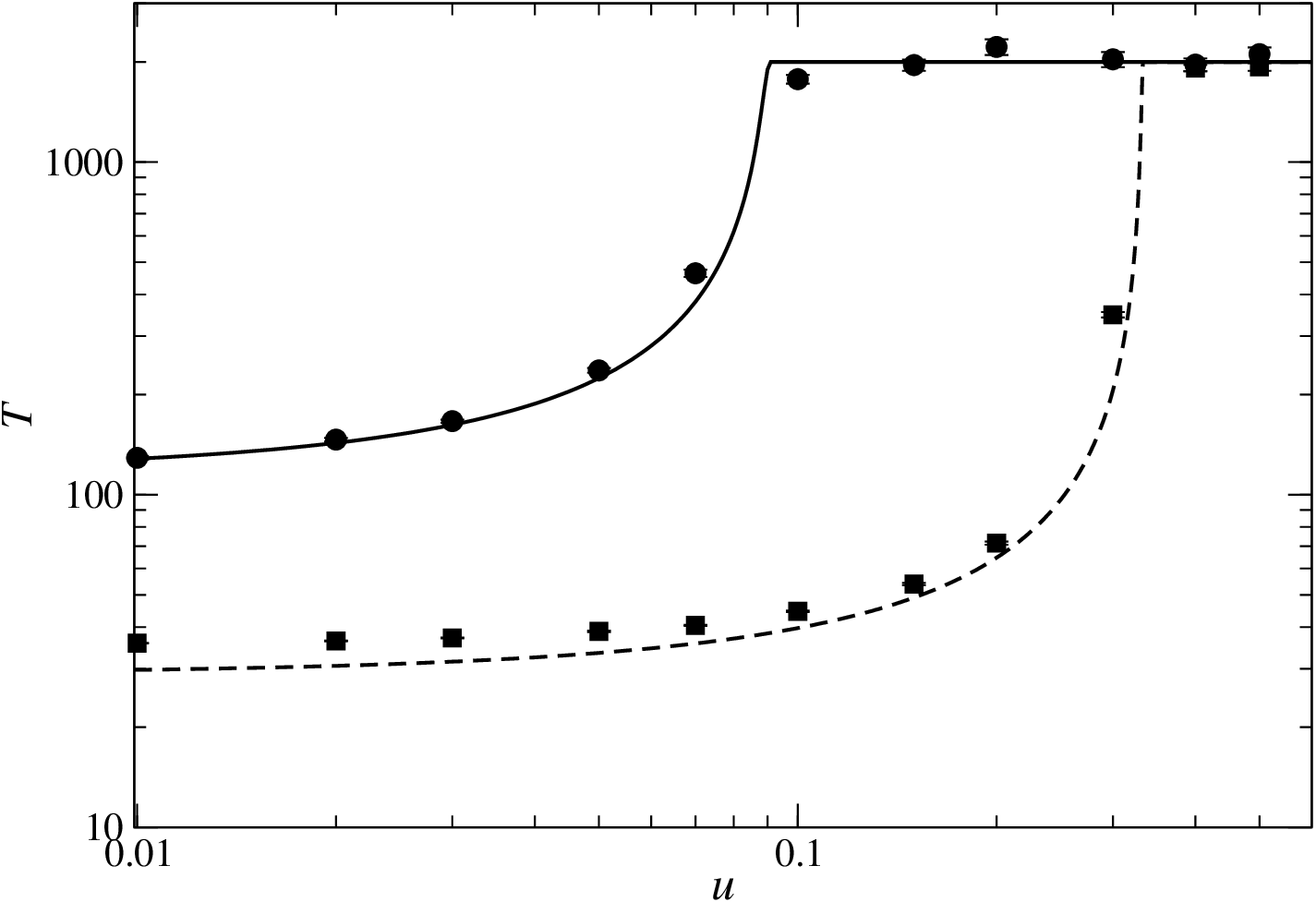}
\caption{\label{figure:fig3} Time to fixation as a function of the mutation
  probability $u$ ($s_{1}=0.1$ and $s_{2}=0.5$, $\Delta t=100$, $N=1000$).
  Solid line: the time $T_{1}$ to fixation for the less beneficial mutant, dashed line:
   time to fixation of more beneficial mutant $T_{2}$.}
\end{figure}

In Figure \ref{figure:fig3} we show the fixation times $T_{1}$ and $T_{2}$ as a
function of mutation rate obtained from simulations, and compared to the prediction
from diffusion theory, Eq. (\ref{Time-Kimura}). As we increase the mutation rate
$u$, the population moves from a strong selection regime, characterized by a short
time to fixation, to a neutral regime, where $T \approx 2N$. We also observe that
the transition between these two regimes occurs at different mutation rates for the
two genotypes: $T_{1}$ shows an abrupt transition around the critical value
$u=u_{1c} \approx 0.09$, whereas $T_{2}$ reaches the neutral regime around
$u=u_{2c} \approx 0.30$. The mutation rate at which this transition occurs is known
as the {\it error threshold}~\citep{eigen}.

\begin{figure}[tbc]
\centering
\includegraphics[bb=14 44 708 527,width=10cm,angle=0,clip=true]{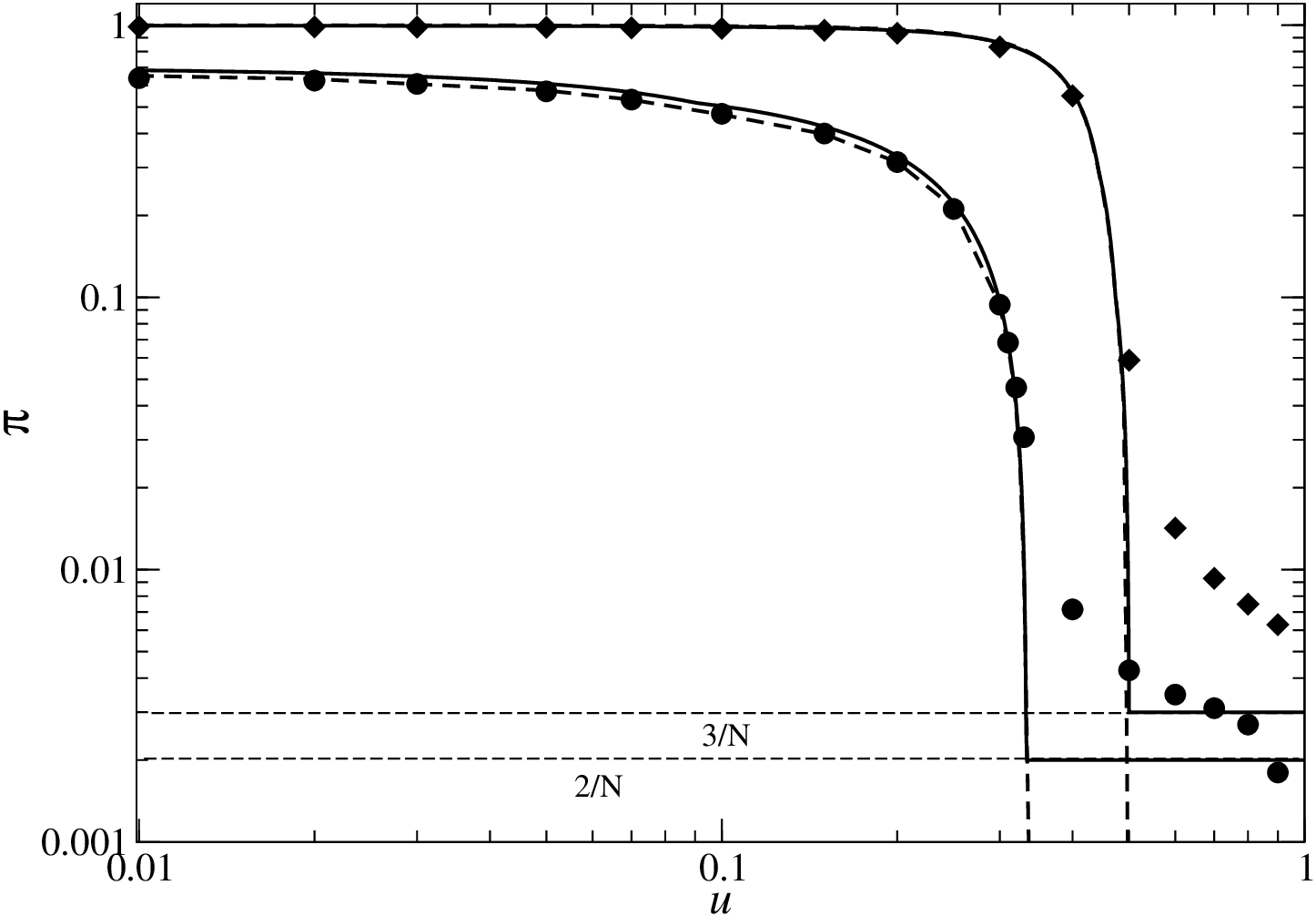}
\caption{\label{figure:fig4} Overall probability of fixation as a function of
  mutation rate $u$. Circles correspond to the situation of two invading
  mutants ($s_{1}=0.1$, $s_{2}=0.5$, $\Delta t=100$, $N=1000$), and squares
  correspond to the situation of three invading mutants ($s_{1}=0.8$, $s_{2}=0.9$,
  $s_3=1.0$, $\Delta t = 20$). The solid line is the prediction according to diffusion theory,
  while the long-dashed line corresponds to the branching process. The short dashes indicate
  the fixation probability levels according to the neutral theory Eq. (\ref{neutral}), for two or three mutants, respectively.}
\end{figure}

In Figure \ref{figure:fig4} we can see the overall probability of fixation as a
function of mutation rate for the case of two as well as three interfering mutants.
Simulation results are again compared to the diffusion theory result Eq.
(\ref{kimura-approx}), but also to a prediction from branching process theory, Eq.
(\ref{branchingmut2}). For these simulations, we used $s_{1}=0.1$, $s_{2}=0.5$ and
population size $N=1000$. We also chose a time interval $\Delta t =100$, which is
smaller than the time required for fixation of mutant $1$, to ensure that the
dynamics takes place in the clonal interference regime. Above the error threshold,
the probability of fixation is known to be $1/N$ for a single mutation ($2/N$ and
$3/N$ for two or three mutants, respectively).  Because the branching process
description assumes infinite population size, it predicts zero probability of
fixation above the error threshold. Diffusion theory, on the other hand, describes
this regime adequately, while being less accurate at small mutation rates. The
abrupt change between the ordered and disordered regime predicted by both theories
is not present in the numerical data because the simulated system is finite.

\begin{figure}[tbc]
\centering
\includegraphics[width=10cm,angle=0,clip=true]{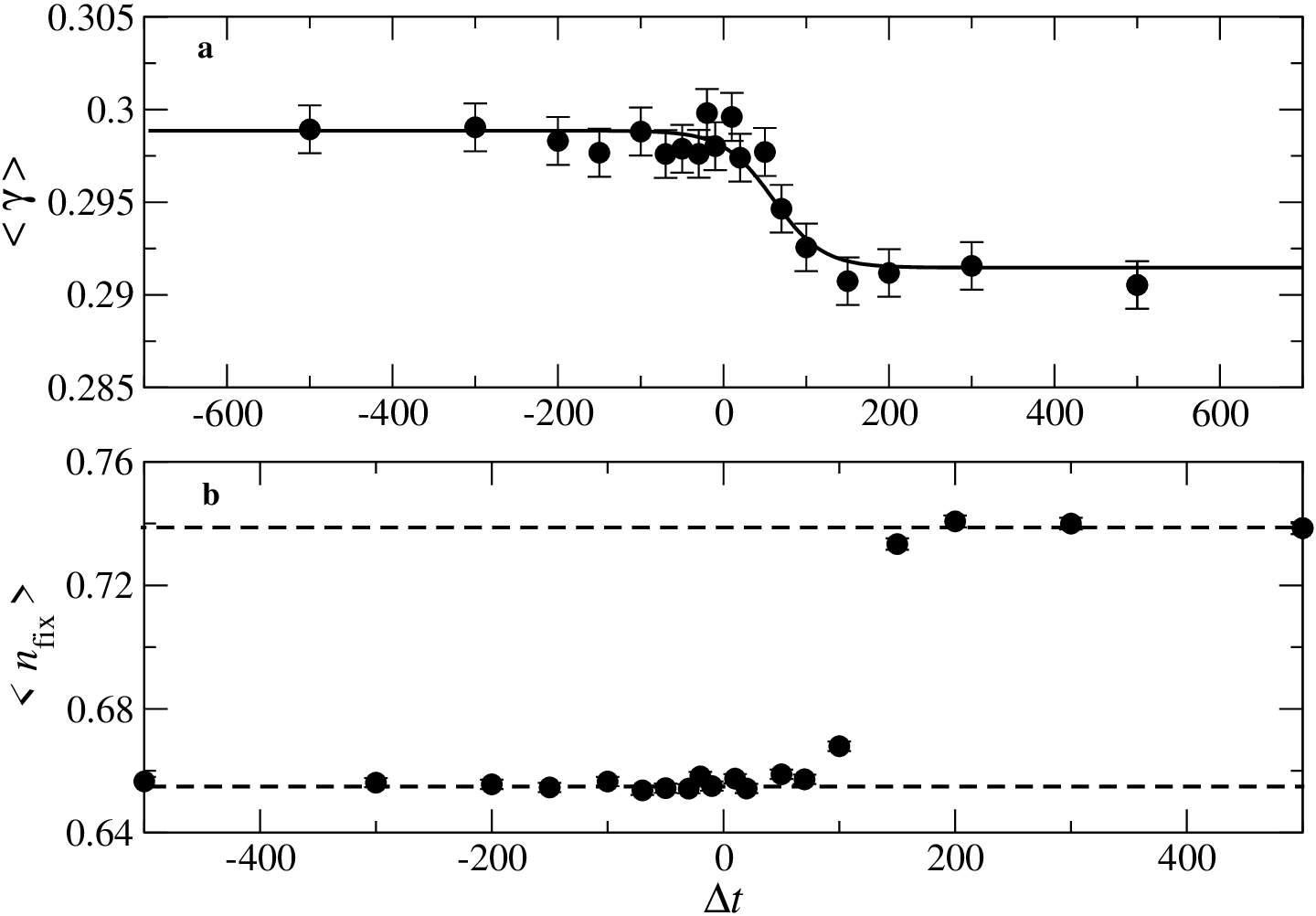}
\caption{\label{figure:figNmutants} a: Expected increase in fitness due to
introduction of mutants as a function of time between their introduction. b:
Expected number of fixed mutants (panel b) as a function of $\Delta t$
($s_{1}=0.1$, $s_{2}=0.5$, $u=0.0$, $N=1000$, for a and batt). Solid line in panel
(a) is expected result according to fixation probabilities given by Eq.
(\ref{tanh-ansatz}).}
\end{figure}

Finally, we tested the prediction that the mean number of fixations in the
population is larger when the mutant with the smaller selective advantage is
introduced first, even though the mean fitness increase is smaller. Figure
\ref{figure:figNmutants}a shows the results for the expected increase in fitness as
a function of the time interval $\Delta t$ as defined in Eq. (\ref{expectedinc}),
as well as the solution of Eq. (\ref{expectedinc}) using a sigmoidal {\it ansatz}
for the probabilities $\pi_{1}$ and $\pi_{2}$, as defined in Eq.
(\ref{tanh-ansatz}). As expected, the mean fitness gain decreases as $\Delta t$
turns positive, while  the expected number of fixed mutants, shown in Figure
\ref{figure:figNmutants}b, increases.

\section{Discussion}
The rate of adaptation of a non-recombining chromosome or asexual population is a
non-trivial quantity that depends on details of the fitness distribution of
mutations. Beneficial mutations often take a long time to dominate in a population,
and compete against each other in the background of deleterious mutations. The
probability of fixation of any one mutation depends on the time of their
introduction, stochastic drift, and the prevalent mutation rate. A good
understanding of the rate of adaptation of asexuals is important in molecular
phylogeny, because it influences the speed of the assumed molecular
clock~\citep{KuhnerFelsenstein1994}. Since successive fixation events determine the
branchings in the phylogenetic tree, a stochastic analysis of the fixation
probability of competing mutations can provide insight into models of evolution
used for tree reconstruction methods~\citep{PageHolmes1998}.

We studied the probability of fixation of beneficial mutations in the presence of
other beneficial mutations that were introduced either earlier or later, and in the
presence of other deleterious mutations, that is, we studied fixation in a
non-equilibrium background. This problem has been addressed previously
by~\cite{barton95} using a deterministic approach, which was unfortunately limited
to small selective advantages. Similarly, Johnson and
Barton~\citeyearpar{johnson2002} studied the probability of fixation in a changing
background, but they confined themselves mostly to the case where deleterious
mutations accumulate in the background after a selective sweep.

We found that the probability of fixation is easily understood from an analytic
point of view as long as the time interval between the introduction of mutants is
larger than the fixation time of the least beneficial mutation. In the interference
regime, where the population consists of clones of the wild type as well as both
beneficial mutants, a phenomenological approach based on logistic growth is
successful at describing the competition. In general, we find that the probability
of fixation of the most beneficial mutant is reduced when it competes against a
less beneficial mutant, and there is a sizable probability that the less beneficial
mutant will in fact survive (see Fig. 1). Yet, the probability that either one of
the mutants survives does not depend on the timing of their introduction.

A number of authors noted the reduction of fixation probabilities caused by
segregating deleterious mutations
\citep{peck94,manning84,charlesworth94,orr2000,johnson2002}, and we observe the
same qualitative behavior here. We find that a background of deleterious mutations
suppresses the fixation probability, and this suppression is more pronounced for
the less beneficial mutant. In the model presented here, there is a simple
relationship between beneficial and deleterious mutations: Each mutation occurring
on a clone created by a beneficial mutation is reduced to the wild type fitness.
This simplified model corresponds to a single-peak landscape, exhaustively studied
in the context of quasispecies theory
\citep{tarazona92,SwetinaSchuster82,galluccio97,camposfonta98}. Consequently,  we
expect a more pronounced decrease of the probability of fixation for small $s$ due
to the existence of an error-threshold, i.e.,  a point at which information is lost
from the sequence due to a critically high mutation rate, see~\cite{eigen}. In the
stochastic regime of evolution, which occurs for mutation rates above the error
threshold, individuals replicate randomly,  and each member of the population is
expected to be the most recent ancestor of the population with the same probability
$\pi=1/N$ while the time to fixation is
$T=2N$~\citep{kimura68,kingman82,donnelly95}. This prediction is confirmed in
Figure \ref{figure:fig3}, where we see that for large $u$ both $T_{1}$ and $T_{2}$
are approximately equal to $2N$. Accordingly, in the interval $u_{1c} < u <
u_{2c}$, where $u_{ic}$ denotes the error-threshold for genotype of type $i$, we
can describe the dynamics by ignoring the less beneficial mutant.

The explicit modeling of stochastic interference between beneficial mutation
revealed an interesting observation concerning the mean number of fixation events
and the expected fitness change. Naively one might surmise that, as the number of
fixation events increases, so does the expected increase in fitness. Instead, we
found the opposite dynamic: Fig. 6 clearly shows that if the mutant with the
smaller selective advantage appears first, then we expect more fixation events
because the second mutant then has a good chance to survive. Yet, this is not the
best scenario if we are interested in maximizing fitness: In this case, it is
better if the most beneficial mutant can establish itself first, even though this
implies that there is little chance for another fixation event.

Finally, we observed that the probability for any beneficial mutant to go to
fixation is equal to the probability that a genotype bearing {\it all} mutations
will become the ancestor of the entire population. Moreover, this observation holds
for an arbitrary number of competing mutations, even in the presence of deleterious
mutations, independently of the order of appearance of the beneficial mutations.
Naturally, while such a probability is very high if the sum of all benefits is
large, the probability of a single sequence appearing with this combination of
mutations is exponentially small in non-recombining populations.  This result was
confirmed using an extension of Kimura's result~\citeyearpar{kimura62} to finite
mutation rates [Eq. (\ref{kimura-approx})] and also by means of a branching process
formulation. Both approaches yield similar results and are in excellent agreement
with our numerical simulations. We find that the branching process approach is more
accurate at low mutation rates, away from the error threshold. As it is formulated
in the $N \rightarrow \infty$ limit, it fails to predict the asymptotic result $\pi
\approx n/N$ which is reached in the limit that all $\gamma_{i}$ vanish.

The present analysis concerns the rate of adaptation as beneficial mutations
compete against each other in the presence of deleterious mutations, in a
non-equilibrium framework. This is a step towards the goal of characterizing the
rate of adaptation of whole populations in arbitrary circumstances, and in the
presence of recombination. We expect that these extensions are necessary before
accurate predictions of optimal mutation rates in biological evolution can be made.

\section*{Acknowledgments}
P.~R.~A. Campos is supported by Funda\c{c}\~ao de Amparo \`a Pesquisa do Estado de
S\~ao Paulo, Proj. No. 99/09644-9. This work is supported by the NSF under contract
No. DEB-9981397. The work of C.A. was carried out in part at the Jet Propulsion
Laboratory (California Institute of Technology), under a contract with the National
Aeronautics and Space Administration.

\bibliographystyle{chicago}
\bibliography{clonal}

\begin{thebibliography}{}

\bibitem[\protect\citeauthoryear{Barton}{Barton}{1995}]{barton95}
Barton, N.~H. (1995).
\newblock Linkage and the limits to natural selection.
\newblock {\em Genetics\/}~{\em 140}, 821--841.

\bibitem[\protect\citeauthoryear{Campos and Fontanari}{Campos and
  Fontanari}{1998}]{camposfonta98}
Campos, P. R.~A. and J.~F. Fontanari (1998).
\newblock Finite-size scaling of the quasispecies model.
\newblock {\em Physical Review E\/}~{\em 58}, 2664--2667.

\bibitem[\protect\citeauthoryear{Charlesworth}{Charlesworth}{1994}]{charleswor%
th94}
Charlesworth, B. (1994).
\newblock The effect of background selection against deleterious mutations on
  weakly selected, linked variants.
\newblock {\em Genet. Res. Camb.\/}~{\em 63}, 213--227.

\bibitem[\protect\citeauthoryear{Cuevas, Elena, and Moya}{Cuevas
  et~al.}{2002}]{cuevas2002}
Cuevas, J.~M., S.~F. Elena, and A.~Moya (2002).
\newblock Molecular basis of adaptive convergence in experimental populations
  of {RNA} viruses.
\newblock {\em Genetics\/}~{\em 162}, 533--542.

\bibitem[\protect\citeauthoryear{de~Visser, Zeyl, Gerrish, Blanchard, and
  Lenski}{de~Visser et~al.}{1999}]{visser99}
de~Visser, J. A. G.~M., C.~W. Zeyl, P.~J. Gerrish, J.~L. Blanchard, and R.~E.
  Lenski (1999).
\newblock Diminishing returns from mutation suply rate in asexual populations.
\newblock {\em Science\/}~{\em 283}, 404--406.

\bibitem[\protect\citeauthoryear{Dobzhansky}{Dobzhansky}{1973}]{Dobzhansky1973}
Dobzhansky, T. (1973).
\newblock Nothing in biology makes sense except in the light of evolution.
\newblock {\em The American Biology Teacher\/}~{\em 35}, 125--129.

\bibitem[\protect\citeauthoryear{Donnelly and Tavar\'e}{Donnelly and
  Tavar\'e}{1995}]{donnelly95}
Donnelly, P. and S.~Tavar\'e (1995).
\newblock Coalescents and genealogical structure under neutrality.
\newblock {\em Annu. Rev. Genet.\/}~{\em 29}, 401--421.

\bibitem[\protect\citeauthoryear{Eigen}{Eigen}{1971}]{eigen}
Eigen, M. (1971).
\newblock Selforganization of matter and evolution of biological
  macromolecules.
\newblock {\em Naturwissenschaften\/}~{\em 58}, 465--429.

\bibitem[\protect\citeauthoryear{Eigen, McCaskill, and Schuster}{Eigen
  et~al.}{1988}]{Eigenetal88}
Eigen, M., J.~McCaskill, and P.~Schuster (1988).
\newblock Molecular quasi-species.
\newblock {\em J. Phys. Chem.\/}~{\em 92}, 6881--6891.

\bibitem[\protect\citeauthoryear{Eigen, McCaskill, and Schuster}{Eigen
  et~al.}{1989}]{Eigenetal89}
Eigen, M., J.~McCaskill, and P.~Schuster (1989).
\newblock The molecular quasi-species.
\newblock {\em Adv. Chem. Phys.\/}~{\em 75}, 149--263.

\bibitem[\protect\citeauthoryear{Ewens}{Ewens}{1979}]{Ewens79}
Ewens, W.~J. (1979).
\newblock {\em Mathematical Population Genetics}.
\newblock Berlin: Springer.

\bibitem[\protect\citeauthoryear{Fisher}{Fisher}{1922}]{fisher22}
Fisher, R.~A. (1922).
\newblock On the dominance ratio.
\newblock {\em Proc. Roy. Soc. Edinb. Sect. B Biol. Sci.\/}~{\em 42}, 321--341.

\bibitem[\protect\citeauthoryear{Fisher}{Fisher}{1930}]{fisher30}
Fisher, R.~A. (1930).
\newblock {\em The Genetical Theory of Natural Selection}.
\newblock Claredon Press.

\bibitem[\protect\citeauthoryear{Galluccio}{Galluccio}{1997}]{galluccio97}
Galluccio, S. (1997).
\newblock Exact solution of the quasispecies model in a sharply peaked fitness
  landscape.
\newblock {\em Phys. Rev. E\/}~{\em 56}, 4526--4539.

\bibitem[\protect\citeauthoryear{Gerrish}{Gerrish}{2001}]{gerrish2001}
Gerrish, P. (2001).
\newblock The rhythm of microbial adaptation.
\newblock {\em Nature\/}~{\em 413}, 299--302.

\bibitem[\protect\citeauthoryear{Gerrish and Lenski}{Gerrish and
  Lenski}{1998}]{gerrish98}
Gerrish, P.~J. and R.~E. Lenski (1998).
\newblock The fate of competing beneficial mutations in an asexual population.
\newblock {\em Genetica\/}~{\em 102/103}, 127--144.

\bibitem[\protect\citeauthoryear{Haldane}{Haldane}{1927}]{haldane27}
Haldane, J. B.~S. (1927).
\newblock A mathematical theory of natural and artificial selection. {Part V}:
  Selection and mutation.
\newblock {\em Proc. Camb. Phil. Soc.\/}~{\em 26}, 220--230.

\bibitem[\protect\citeauthoryear{Harris}{Harris}{1963}]{harris63}
Harris, T.~E. (1963).
\newblock {\em The Theory of Branching Processes}.
\newblock Springer.

\bibitem[\protect\citeauthoryear{Hill and Robertson}{Hill and
  Robertson}{1966}]{hill66}
Hill, W.~G. and A.~Robertson (1966).
\newblock The effect of linkage on the limits to artificial selection.
\newblock {\em Genet. Res.\/}~{\em 8}, 269--294.

\bibitem[\protect\citeauthoryear{Johnson and Barton}{Johnson and
  Barton}{2002}]{johnson2002}
Johnson, T. and N.~H. Barton (2002).
\newblock The effect of deleterious alleles on adaptation in asexual organisms.
\newblock {\em Genetics\/}~{\em 162}, 395--411.

\bibitem[\protect\citeauthoryear{Kimura}{Kimura}{1962}]{kimura62}
Kimura, M. (1962).
\newblock On the probability of fixation of mutant genes in a population.
\newblock {\em Genetics\/}~{\em 47}, 713--719.

\bibitem[\protect\citeauthoryear{Kimura}{Kimura}{1968}]{kimura68}
Kimura, M. (1968).
\newblock Evolutionary rate at the molecular level.
\newblock {\em Nature\/}~{\em 217}, 624--626.

\bibitem[\protect\citeauthoryear{Kimura and Ohta}{Kimura and
  Ohta}{1969}]{KimuraOhta69}
Kimura, M. and T.~Ohta (1969).
\newblock The average number of generations until fixation of a mutant gene in
  a finite population.
\newblock {\em Genetics\/}~{\em 61}, 763--771.

\bibitem[\protect\citeauthoryear{Kingman}{Kingman}{1982}]{kingman82}
Kingman, J. F.~C. (1982).
\newblock On the genealogies of large populations.
\newblock {\em J. Appl. Prob.\/}~{\em 19A}, 27--43.

\bibitem[\protect\citeauthoryear{Kuhner and Felsenstein}{Kuhner and
  Felsenstein}{1994}]{KuhnerFelsenstein1994}
Kuhner, M. and J.~Felsenstein (1994).
\newblock A simulation comparison of phylogeny algorithms under equal and
  unequal evolutionary rates.
\newblock {\em Mol. Biol. Evol.\/}~{\em 11}, 459--68.

\bibitem[\protect\citeauthoryear{Manning and Thompson}{Manning and
  Thompson}{1984}]{manning84}
Manning, J.~T. and D.~J. Thompson (1984).
\newblock Muller's ratchet and the accumulation of favourable mutations.
\newblock {\em Acta Biotheor.\/}~{\em 33}, 219--225.

\bibitem[\protect\citeauthoryear{McVean and Charlesworth}{McVean and
  Charlesworth}{2000}]{McVeanCharlesworth2000}
McVean, G. A.~T. and B.~Charlesworth (2000).
\newblock The effects of {Hill-Robertson} interference between weakly selected
  mutations on patterns of molecular evolution and variation.
\newblock {\em Genetics\/}~{\em 155}, 929--944.

\bibitem[\protect\citeauthoryear{Miralles, Gerrish, Moya, and Elena}{Miralles
  et~al.}{1999}]{miralles99}
Miralles, R., P.~J. Gerrish, A.~Moya, and S.~F. Elena (1999).
\newblock Clonal interference and the evolution of {RNA} viruses.
\newblock {\em Science\/}~{\em 285}, 1745--1747.

\bibitem[\protect\citeauthoryear{Muller}{Muller}{1964}]{Muller64}
Muller, H.~J. (1964).
\newblock The relation of recombination to mutational advance.
\newblock {\em Mutat. Res.\/}~{\em 1}, 2--9.

\bibitem[\protect\citeauthoryear{Orr}{Orr}{2000}]{orr2000}
Orr, H.~A. (2000).
\newblock The rate of adaptation in asexuals.
\newblock {\em Genetics\/}~{\em 155}, 961--968.

\bibitem[\protect\citeauthoryear{Page and Holmes}{Page and
  Holmes}{1998}]{PageHolmes1998}
Page, R. and L.~Holmes (1998).
\newblock {\em Molecular Evolution: A Phylogenetic Approach}.
\newblock Blackwell Science.

\bibitem[\protect\citeauthoryear{Peck}{Peck}{1994}]{peck94}
Peck, J.~R. (1994).
\newblock A ruby in the rubbish: Beneficial mutations, deleterious mutations
  and the evolution of sex.
\newblock {\em Genetics\/}~{\em 137}, 597--606.

\bibitem[\protect\citeauthoryear{Rozen, de~Visser, and Gerrish}{Rozen
  et~al.}{2002}]{rozen2002}
Rozen, D.~E., J.~A. G.~M. de~Visser, and P.~J. Gerrish (2002).
\newblock Fitness effects of fixed beneficial mutations in microbial
  populations.
\newblock {\em Current Biology\/}~{\em 12}, 1040--1045.

\bibitem[\protect\citeauthoryear{Shaver, Dombrowski, Sweeney, Treis, Zappala,
  and Sniegowski}{Shaver et~al.}{2002}]{shaver2002}
Shaver, A.~C., P.~G. Dombrowski, J.~Y. Sweeney, T.~Treis, R.~M. Zappala, and
  P.~D. Sniegowski (2002).
\newblock Fitness evolution and the rise of mutator alleles in experimental
  {{\it Escherichia coli}} populations.
\newblock {\em Genetics\/}~{\em 162}, 557--566.

\bibitem[\protect\citeauthoryear{Swetina and Schuster}{Swetina and
  Schuster}{1982}]{SwetinaSchuster82}
Swetina, J. and P.~Schuster (1982).
\newblock Self-replication with errors: {A} model for polynucleotide
  replication.
\newblock {\em Biophys. Chem.\/}~{\em 16}, 329--345.

\bibitem[\protect\citeauthoryear{Tarazona}{Tarazona}{1992}]{tarazona92}
Tarazona, P. (1992).
\newblock Error thresholds for molecular quasi-species as phase-transitions -
  from simple landscapes to spin-glass models.
\newblock {\em Physical Review A\/}~{\em 45}, 6038--6050.

\bibitem[\protect\citeauthoryear{van Nimwegen, Crutchfield, and Mitchell}{van
  Nimwegen et~al.}{1999}]{vanNimwegenetal99a}
van Nimwegen, E., J.~P. Crutchfield, and M.~Mitchell (1999).
\newblock Statistical dynamics of the {Royal Road} genetic algorithm.
\newblock {\em Theoretical Computer Science\/}~{\em 229}, 41--102.

\bibitem[\protect\citeauthoryear{Wilke}{Wilke}{2003}]{wilke2003}
Wilke, C.~O. (2003).
\newblock Probability of fixation of an advantageous mutant in a viral
  quasispecies.
\newblock {\em Genetics\/}~{\em 162}, 467--474.

\bibitem[\protect\citeauthoryear{Wilke, Ronnewinkel, and Martinetz}{Wilke
  et~al.}{2001}]{Wilkeetal2001a}
Wilke, C.~O., C.~Ronnewinkel, and T.~Martinetz (2001).
\newblock Dynamic fitness landscapes in molecular evolution.
\newblock {\em Phys. Rep.\/}~{\em 349}, 395--446.

\end{thebibliography}

\begin{appendix}
\renewcommand{\theequation}{\thesection.\arabic{equation}}
\setcounter{equation}{0}
\section{Probability of fixation and time to fixation for an individual mutant}

We can calculate the probability of fixation of an individual mutant either from
branching process theory
\citep{fisher22,haldane27,fisher30,barton95,johnson2002,wilke2003} or from
diffusion theory \citep{kimura62}.  Branching process theory is applicable for very
large population sizes, and arbitrary (but positive) selective advantages $s$.
Diffusion theory is applicable to moderately large to large population sizes and
small or even vanishing selective advantages $s$. Moreover, diffusion theory also
gives an expression for the expected time to fixation.

\subsection{Diffusion theory}

If the advantageous mutant suffers additional mutations while it goes to fixation,
then we typically have to use multi-dimensional diffusion equations, which can be
very unwieldy. However, in the simple case treating the advantageous mutant and the
wild type only, and where the only effect of deleterious mutations is to revert the
advantageous mutant back to wild type, one-dimensional diffusion theory, as
developed by \citet{kimura62,KimuraOhta69}, is still applicable, and gives
satisfying results \citep{vanNimwegenetal99a,Wilkeetal2001a}.

The main quantities in diffusion theory are the mean $M_{\delta x}$ and the
variance $V_{\delta x}$ in the rate of change per generation of the
concentration $x$ of the invading mutant. The derivation of
these quantities is straightforward \citep{Ewens79}, and we find
\begin{align}
  M_{\delta x}&= \gamma x (1-x)\,,\\
V_{\delta x}&=x(1-x)/N\,
\end{align}
to first order in $1/N$ and $\gamma$. Here, $\gamma$ is the relative difference in
mean fitness between a wild-type population and a population with fixed
advantageous mutant. The only difference between these expressions and those of
\citet{kimura62} is that $\gamma$ replaces the selective advantage $s$ in
$M_{\delta x}$. Without mutations ($u=0$), $\gamma$ is equal to $s$, and we recover
the standard results. For positive $u$, $\gamma$ has the form
\begin{equation}\label{gamma-def}
  \gamma = \left\{\begin{matrix} (1+s)(1-u) -1 & \mbox{for $u<s/(1+s)$,} \\ 0 &
  \mbox{for $u\geq s/(1+s)$.} \end{matrix}\right.
\end{equation}
The mutation rate $u_{\rm c}$ at which $\gamma$ reaches 0, %
$u_{\rm c}=s/(1+s)$, corresponds to the error threshold of quasispecies theory
\citep{SwetinaSchuster82,Eigenetal88,Eigenetal89}. For mutation rates larger than
$u_{\rm c}$, the invading mutant does not have an advantage over the wild type, and
the fixation process corresponds to that of a neutral mutant.

The probability of fixation follows now from \citet{kimura62} as
\begin{equation}\label{kimura-approx}
  P(s,u) = \frac{1-e^{-2\gamma}}{1-e^{-2\gamma N}}\,.
\end{equation}
Likewise, the expected time to fixation follows from \citet{KimuraOhta69} as
\begin{equation}\label{Time-Kimura}
  T(s,u) = J_1 + \frac{1-P(s,u)}{P(s,u)}J_2\,,
\end{equation}
where
\begin{align}
  J_1 & = \frac{1}{\gamma (1-e^{-2\gamma N})} \int_{1/N}^1
    \frac{(e^{2\gamma N x}-1)(e^{-2\gamma N x}-e^{-2\gamma N})}{x(1-x)}dx\,,\\
  J_2 & = \frac{1}{\gamma (1-e^{-2\gamma N})}\int_0^{1/N}
    \frac{(e^{2\gamma N x}-1)(1-e^{-2\gamma N x})}{x(1-x)}dx\,.
\end{align}

\subsection{Branching process theory}

According to \citet{barton95}, the probability that a beneficial mutation reaches
fixation in a genotype with genetic background $i$ follows from iterating the
following set of equations:
\begin{equation}\label{iterative}
(1-P_{i,t-1})=\sum_{j=0}^{\infty}W_{i,j}(1-P_{i,t}^{*})^{j}\,,
\end{equation}
where $P_{i,t}$ is the probability of fixation of an allele that is present in a
single copy in site $i$ in generation $t$, $W_{i,j}$ denotes the probability that
an allele in site $i$ contributes with  $j$ offsprings to the next generation, and
\begin{equation}
P_{i,t}^{*}=\sum_{k}M_{i,k}P_{k,t}
\end{equation}
is the probability that an allele in background $i$ at time $t-1$ would be
fixed, given that at time {\it t} it is passed to one offspring. The quantity
$M_{j,k}$ gives the probability that an offspring from a parent at background
$i$ will be at background $k$.  If the distribution of offspring is given by a
Poisson distribution with mean $(1+s_{i})$, i.e.,
\begin{equation}
W_{i,j} = \frac{(1+s_{i})^{j}}{j!}e^{-(1+s_{i})},
\end{equation}
then Eq. (\ref{iterative}) becomes
\begin{equation}\label{iterative1}
(1-P_{i,t-1})=\exp \left[-(1+s_{i})P_{i,t}^{*} \right].
\end{equation}
The fixation probabilities correspond to the solution of Eq. (\ref{iterative})
obtained in the limit $t \rightarrow \infty$.

 In our model, we consider only
two distinct classes of genotypes, and mutations occur only from advantageous
mutant to wild type.  Because the wild type has a mean number of offspring per
generation exactly equal to $1$, the probability of fixation of wild type
sequences in the branching-process formulation vanishes \citep{harris63}.
Therefore, the probability of fixation $P(s,u)$ is given by the solution of the
equation:
\begin{equation}\label{branchingmut}
P(s,u)=1-\exp\left[-(1+s)(1-u)P(s,u)\right].
\end{equation}
Unfortunately, it is not possible to derive a closed-form solution to this
equation. $P(s,u)$ has to be determined numerically from iterating
Eq.~(\ref{branchingmut}).

Since the relevant parameter seems to be $\gamma=(1+s)(1-u)-1$, the above equation
can be written as
\begin{equation}\label{branchingmut1}
P(s,u)=1-\exp\left[-(\gamma+1)P(s,u)\right].
\end{equation}
Although a derivation can not be performed when $n$ beneficial mutants are considered, we
found empirically that in this case
\begin{equation}\label{branchingmut2}
P(s,u)=1-\exp\left[-\left[\prod_{i=1}^{n}(\gamma_{i}+1)\right]P(s,u)\right].
\end{equation}
This expression is used to compare numerical results to the branching process
theory prediction in Fig.~\ref{figure:fig4}.
\end{appendix}
\end{document}